\documentclass[10pt, conference, compsocconf]{IEEEtran}
\usepackage{graphicx}
\usepackage{comment}
\usepackage{amsmath}
\usepackage{amsfonts}
\usepackage{url}
\usepackage{xspace}
\usepackage{fancyvrb}

\title{Secure Abstraction with Code Capabilities}

\author{\IEEEauthorblockN{Robbert van Renesse}
\IEEEauthorblockA{Cornell University\\
rvr@cs.cornell.edu}
\and
\IEEEauthorblockN{H{\aa}vard Johansen}
\IEEEauthorblockA{University of Troms{\o}\\
haavardj@cs.uit.no}
\and
\IEEEauthorblockN{Nihar Naigaonkar}
\IEEEauthorblockA{Cornell University\\
niharnaigaonkar@gmail.com}
\and
\IEEEauthorblockN{Dag Johansen}
\IEEEauthorblockA{University of Troms{\o}\\
dag@cs.uit.no}
}

\newcommand{\pr}[1]{\langle #1 \rangle}

\begin{document}

\maketitle

\begin{abstract}
We propose embedding executable code fragments in cryptographically protected capabilities to
enable flexible discretionary access control in cloud-like computing infrastructures.
We are developing this as part of a sports analytics application that runs on a federation of public and enterprise clouds.
The capability mechanism is implemented completely in user space.
Using a novel combination of X.509 certificates and Javscript code, the capabilities support restricted delegation,
confinement, revocation, and rights amplification for secure abstraction.
\end{abstract}

\section{Introduction}

The predominant way of providing discretionary access control in
the cloud is through a combination of authentication and access
control lists.  But such mechanisms are not without problems.  People
and even entire companies end up with accounts in many different places.
While single-signon mechanisms exist, they are adopted sparingly.
To deal with their many accounts, people often use the
same user name and password everywhere, or variations on a password
that are easy to generate (\emph{e.g.}, ``mypwd4amazon''), but also
easy to reverse engineer.  A malicious administrator at one site can then
access accounts of users at other sites.

This problem with access control list became apparent while developing 
Muithu~\cite{Joha1208:Muithu}, a sports analytics application that runs on a federation of public and enterprise clouds.  
A wealth of performance data is being collected in real-time by teams, sports media, and spectators.
Careful analysis of such data is a crucial part of competitive sports.
Much of the data is private and highly sensitive;
this includes medical performance data, internal individual performance evaluations, and future training strategies.

An important part of Muithu is abstraction.  Raw data from various
sources are processed and made available in another form, 
so that multiple layers of abstraction can be developed.
With access control lists, each layer would need to have
accounts with the lower layers, and also keep track of accounts
of its own users and manage who is allowed to access which data.
Much of the complexity then revolves around securely managing
user accounts and correctly configuring the access control lists.
Access control lists make it difficult to maintain fine grained
control over distribution and access of data.

The mechanism we propose here does not require authenticating any
users because authorization is done through \emph{capabilities}.
Capabilities are unforgeable digital tokens that can be passed
around, and possession of a capability grants specific rights to
services independent of who the possessor is.  Consistent with the
Principle of Least Privilege, capabilities are given out on an
as-needed basis.  Capabilities have been used in a variety of
systems.  The instantiation of capabilities that we propose is novel
in a variety of ways:

\begin{itemize}
\item the capabilities contain embedded code that allow fine-grained
control over restricted delegation.  In other words, the set of rights
that can be delegated is not predefined as in most capability-based
systems but can be evolved as needed;
\item to support secure abstraction, the proposed capabilities support
rights amplification;
\item no special trusted language, trusted operating system kernel,
or other trusted infrastructure is required---the capabilities are
managed completely in user space using public key cryptographic techniques;
\item even though managed in user space, transfer of capabilities is
implicitly mediated so that confinement can be supported;
\item a directory service provides a secure way for users to manage
their capabilities, and to delegate restricted capabilities to other
users.
\end{itemize}

\noindent
We call these capabilities ``code capabilities'' or \emph{codecaps}
for short.

\section{Secure Abstraction}

\emph{Notational Analytics} has become a competitive advantage for
many elite sport coaches resulting in an emerging sport analytics
industry.  Example data include physical variables of individual
athletes like speed, distance covered, agility, energy consumption,
and muscle force.  Such objective physical data is acquired using
body-area sensors and from vision algorithms parsing video feeds.
Additional data is added by expert analysts like whether a soccer pass
was successful or not and how well a team is performing.  Major team
sports like baseball, basketball, and soccer are avid users of such
analytics systems. 

In close collaboration with a Norwegian major-league soccer club,
we developed Muithu, a cloud-based notational analytics system for
recording and analyzing soccer team performance data.  A key requirement
for Muithu was the ability to externalize collected data to third
parties that, for instance, specialize in complex sports analytics.
Also, a recent trend is to publish performance data on social media
and more traditional broadcasting channels.  A new generation of sport
viewers familiar with social networks and micro-blogs tend to prefer this
type of information while watching sport events.  For instance, during
the last European soccer championship in June 2012, major broadcasters
distributed real-time performance data on social media platforms and
traditional television broadcasts while games unfolded.  This included
statistics about successful passes, number of corners, attempted shots
on goal, meters covered by individual players and the like.
Obviously, there are strong security constraints related to athlete and
team performance data.  In particular, medical related information like
heart-rate and injuries are highly personal and cannot be
made public.

The architecture of Muithu is designed to simplify the development of
new sports analytics applications while observing security requirements
from the ground up.  One can think of Muithu as consisting of layers
of abstraction.  Each layer implements its own services and supports
operations through a remote procedure call mechanism.  Access to data
is mediated through codecaps.  Services are run by principals; clients
that access services are principals as well.

The base-layer of Muithu consist of captured notational data, video
feeds, and sensor data that are pushed to and stored on an enterprise
cloud platform through a REST API.  This set of data, hosted by the
base-layer principal $P_0$, is represented as a set of data objects that
can be accessed through a simple interface.  Such data objects may, for
instance, correspond to raw sensor data of individual players in the team,
and might be updated as new data about that player becomes available.
Additional layers are then added as the data is being processed and
tagged.  Some layers have significant cloud resources available, but
others work more like a library executed by their clients, often using
JavaScript in the browser.  The cloud resources of such layers are
only accessed when the library cannot handle requests itself.

As an example, consider the situation where a team coach $P_1$ wants to
provide up-to-date information about each player object $o$ to the local
supporter club $P_2$.  However, $P_1$ has no interest in running a large
web site to share this information.  Instead, $P_1$ can obtain a codecap
$c_1$ from $P_0$ for $o$ and give $P_2$ a library and a delegated codecap
$c_2$ for $o$.  When $P_2$ invokes the library, the library can use $c_2$
to access the current version of $o$ directly from $P_0$ and generate
the derived object $o'$ using the client's computational resources. 
Code in $c_2$ ensures that $P_2$ can only access those parts of $o$ that
$P_1$ allows it to access.

\begin{figure}
\includegraphics[width=\columnwidth]{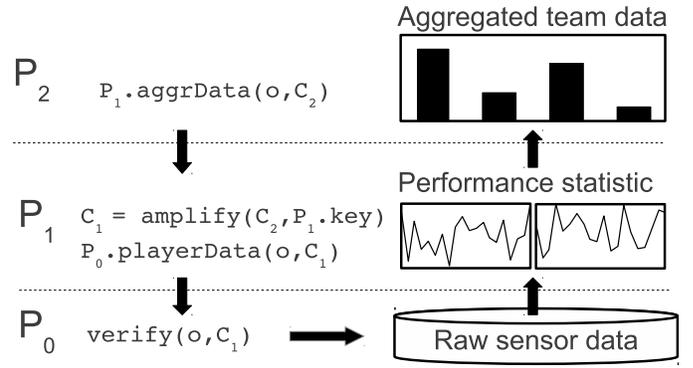}
\caption{Muithu data layering example}
\label{fig:muithu}
\end{figure}

Now suppose that there are certain proprietary operations on $o$ that
$P_1$ does not want to distribute in the library itself or using parts
of the data in $o$ that $P_1$ does not want $P_2$ to access directly.
For instance, $P_1$ might not want to give access to detailed heart-rate
information, but instead provide only access to aggregated values.
In that case the library can accesses a service run by $P_1$ to execute
the operation using codecap $c_2$, as illustrated in Figure~\ref{fig:muithu}.
  $P_1$ cannot use $c_2$ directly
to access $o$ because it does not have the corresponding private key
and because it does not give the necessary access rights.  However,
as we shall see, $P_1$ can reconstruct $c_1$ from $c_2$ and pair the
resulting code cap with its own private key to obtain the correct access
credentials to $o$.  This is a case of \emph{rights amplification},
a necessary ingredient of secure abstraction.  It is not necessary
for $P_1$ to keep around all the intermediate codecaps, which would be
inconvenient and waste computing resources.

In general, abstraction often involves more than one object, and
consequently more than one codecap.  When a client requests an
object, it obtains both a library for the object and the collection
of codecaps that the library needs to access the underlying data
for the objects.  Only clients need to keep track of codecaps, as
rights amplification allows the lower layers to reconstruct them
as necessary.  This much simplifies building secure cloud services
compared to one based on access control lists in which user accounts
must be managed and credentials for lower layers must be stored.

\section{Code Capabilities}

The implementation of codecaps is based on standard
certificate chains.
Each principal $P$
is identified by its public key $P.\texttt{pubkey}$
and has a corresponding private key $P.\texttt{privkey}$
that it keeps carefully hidden from other principals.
In order for a client
to execute a request as some service, the client needs
a codecap for the request.

A codecap $c_n$ is a pair $\pr{h_n, k_n}$ consisting of a
\emph{heritage} and a \emph{private key}.
The heritage $h_n$ is a chain of public
key certificates $[ C_1 :: C_2 :: ... :: C_n ]$ corresponding to a
chain of $n+1$ principals $P_0 ... P_n$.
(The operator $::$ denotes list concatenation.)
In this case, $P_0$ has delegated certain rights to $P_1$, $P_1$,
has delegated rights to $P_2$, ..., and $P_{n-1}$ has
delegated rights to $P_n$.
Certificate $C_i$ is
signed by $k_{i-1} = P_{i-1}.\texttt{privkey}$.
$k_n$ is the private key of $P_n$.
Codecap $c_n$ is owned by principal $P_n$ and gives access
rights to services provided by principal $P_0$.
However, $P_0$ does not have access control lists, does not need to
know anything about $P_n$, and only needs to maintain its private
key $k_0$.

Each certificate $C_i$ is a collection of \emph{attributes} signed
by a private key.  An
attribute is a pair consisting of a name and a value.
We denote by $C_i.\texttt{attr}$ the value of the
attribute named ``\textit{attr}'' in certificate $C_i$.
Each certificate $C_i$ has at least the following attributes:

\begin{itemize}
\item $C_i.\texttt{pubkey}$: contains $P_i.\texttt{pubkey}$;
\item $C_i.\texttt{rights}$:
contains a boolean function that
takes a request as argument and returns \texttt{true} iff the
function allows the request.
\end{itemize}

\noindent
Note that the validity of a heritage can be checked by anybody
who knows $P_0.\texttt{pubkey}$, and that the private key $k_n$
in the codecap is the private key corresponding to the last
certificate $C_n$ on the heritage.
A request is itself a certificate, signed by $k_n = P_n.\texttt{privkey}$.
In some sense the request is appended to the end of the heritage as
a certificate $C_{n+1}$ as if delegated.
The attributes in the request describe the request type and its various
parameters.
Principal $P_0$ will execute the request only if
heritage $h_n$ is valid, the request's signature can be verified,
and if $C_i.\texttt{rights}(r)$ holds for all $i$ in $1 ... n$.

Principal $P_0$ determines the programming language in which the
rights functions are expressed.  The language can be very simple.
For example, a file service might have a language that consists of
only three programs: ``R'', ``W'', and ``RW''.
When the program ``R'' is applied to an update operation, it evaluates
to \texttt{false}.

We intend the language to be Turing-complete and to provide
powerful library functions, such as JavaScript.
For example, say that a file service only provides ``read'' and ``write''
operations and we want to create a codecap that can ``increment''
an integer that is stored in the file.  The client would first read
the file and then write back the incremented value.  The rights
function in the codecap would check that the value that is to be
written is an integer that is one higher than the integer stored
in the file.  Rights functions may also be able to read the clock
on the server.  This can be used to implement expiration times on
codecaps, or, for example, to specify that an operation is
only allowed during daytime.

It is important that such rights functions cannot have
external effects (such as writing files or sending messages) and
that the functions have finite running times.  They must be carefully
sandboxed; loops and recursion may be disallowed and running times
may be limited by a timer.

\section{Using CodeCaps}

To illustrate how codecaps are used, 
suppose a client $P_n$ has a codecap $c_n$ for a service
provided by $P_0$ and wants $P_0$ to execute a request $r$.
To do so, client $P_n$ sends a message $m$ to $P_0$ that contains the
following attributes:
\begin{itemize}
\item $m.\texttt{request}$: a certificate that described the requested
operation and is signed by $P_n.\texttt{privkey}$;
\item $m.\texttt{heritage}$: contains $h_n$, the heritage of the
codecap needed to execute the request.
\end{itemize}

Upon receipt of a message $m$, $P_0$
verifies the heritage, and verifies the signature on the request
certificate using $C_n.\texttt{pubkey}$.
$P_0$ then checks that all rights functions
$C_i.\texttt{rights}(m.\texttt{request})$ return \texttt{true}.
For example, a rights function might express
$m.\texttt{request}.\texttt{type} = \mbox{\sc READ} \wedge
m.\texttt{request}.\texttt{offset} \ge 256$.
If verified, 
$P_0$
executes $m.\texttt{request}$
and returns the result to client $P_n$.

Note that an eavesdropper on the network may intercept the request
message and obtain the heritage of the codecap.  However, without the
corresponding private key, the eavesdropper will not be able to
sign new requests with it.
The eavesdropper can replay the request---it is thus important that
either the service is capable
of eliminating duplicates or that requests are idempotent.
In practice, communication between a client and a service is
usually over SSL, eliminating this concern.

There are two ways in which a codecap can be created.
The first is from scratch, when a new service is offered or a new
client is added. The
second is by (possibly restricted) delegation, in which
case a client communicates one of its codecaps to another principal.
Note that only heritages of codecaps are communicated between
principals---the recipient of the heritage of a new codecap has to 
complete the codecap by pairing it with its private key.

We illustrate here how \emph{confinement} can be achieved. 
A principal $P_n$ can create a codecap for $P_{n+1}$ so that
$P_{n+1}$ cannot delegate rights of that codecap to other
principals without revealing its private key to those principals.
The idea is that the rights function in certificate $C_{n+1}$ has the
ability to test if it is the rights function of the last certificate
in the heritage of the codecap, returning \texttt{false} if not.
If $P_{n+1}$ is faulty it can share its private key with other
principals, but this does not extend the damage from having delegated to
$P_{n+1}$ in the first place.

When confined, a principal $P_n$ that wants to delegate to a principal
$P'_n$ must ask one of the principals on the heritage of the codecap
to generate a codecap for $P'_n$.  In the limit, a service may choose
to confine all its codecaps and thus be involved whenever delegation
takes place.

\section{CodeCap Directories}

Clients and services may end up owning many codecaps.
All codecaps of a principal have the same private key, which the
principal has to maintain securely.
To simplify management of all the heritages and delegation,
we are developing a distributed directory service.
Directories are objects that map string names to codecaps.
However, different from ordinary directory services, a ``lookup''
operation is a restricted delegation:
the directory service delegates its rights to its client.

A directory has rows and columns.  Both rows and columns have names.
There are no two rows with the same name, and no two columns with
the same name.  The first column is called ``name'' and contains
the name of the row.  The second column is called ``cap'' and
contains the heritage of a codecap in each row. The remaining columns
contain rights functions.  Each such column is called a \emph{group}.
Directories support an operation ``chmod'' by which rights functions
in the group columns may be updated.
The execution of the chmod operation itself is restricted by rights
expressed in the directory codecap.

A directory codecap gives access to one or more groups within a directory.
Given a directory codecap $\textit{dc}$,
the operation $\textit{lookup}(\textit{dc}, \textit{name}, \textit{group})$
first finds the row for the given \textit{name}.
In the row it retrieves a heritage $h_n$ in the ``cap'' column and the rights
function $R$ in the given group.
The directory service then delegates its rights given by $h_n$ by
appending a new heritage $h_{n+1}$ using $R$ and signed by the private
key of the directory service.
The directory service then returns the result to the
client, which uses $h_{n+1}$ and its private key to construct a codecap.

Since directories are objects themselves, they may be organized in
any arbitrary directed graph structure (it does not have to be a tree
and can contain cycles).  A user then needs to hold
only one codecap, that of its ``home directory''. Given the codecap
of its home directory, all objects reachable from that directory,
subject to the restrictions specified in the rights functions, are
accessible to the user.  Note that it is not necessary that all
directories are serviced by the same physical server.  In a large
scale system there may be many directory servers in different
geographical locations.

We do not run public directory services, however, as this would
be tantamount to simulating access control lists using codecaps.
Directories are privately owned by principals and run by those
principals to keep track of their own codecaps and to help
with delegating codecaps to other users.

The directory service library supports path names of the form
``/a/b/c''.  The library maintains two directories: that of the
home directory and that of the working directory.  Path names that
start with ``/'' are evaluated relative to the home directory while
other path names are evaluated relative to the working directory.
Initially the home directory and the working directory are the same.
A library ``chdir'' method updates the working directory.


\section{Revocation}

The ``chmod'' operation (as well as the ``remove'' operation) on
directories provide a means to do selective revocation, preventing users
from obtaining codecaps.  However, codecaps
that have already been distributed remain valid.
Various ways have been proposed to revoke outstanding capabilities.
(For an early approach, see~\cite{Red74}.)
One is to associate version numbers with objects~\cite{GGT97}.
A codecap would be for a version of the object,
and certificate $C_1$ would contain the version number the codecap refers
to.
When a service wants to invalidate outstanding codecaps on one of
its objects, it simply increments the version number of the object.
(This technique may also be used for key rotation or dealing with
lost private keys.)

This only works for the raw objects.  If an intermediate service
wants to revoke delegated codecaps, it must ask the provider of
the raw object to increment the version number.  Selective revocation
can be supported with this scheme by having multiple version numbers
per object, that is, one version number for each group of principals.
Alternatively, services can build expiration times into the
rights functions of codecaps as described above.  Clients should
think of such codecaps as ``soft references'' that may at any time
become invalid.
Those clients should be prepared to acquire new codecaps when
necessary.

Another revocation technique exploits indirection.  An intermediate service,
instead of passing out delegated codecaps, could generate fresh codecaps
and act as a proxy to the service that provides the raw objects.
Such a scheme also supports selective revocation
in which only a subset of clients are affected.  
This proxy scheme complicates the intermediate service (in a similar
way as maintaining access control lists) and consequently
has security disadvantages compared to the simple scheme of revoking
all outstanding codecaps.  Whether to use one scheme or another can be
determined by each application individually.

A weakness of codecaps compared to access control lists is that there is no
way to review which principals have rights to a service~\cite{Gli79}.
One option is for a service to confine all its codecaps so it is
involved in and can keep track of all delegation.

\section{Object Lifetimes}

So far we have only considered operations on objects (and services)
that already exist.
We now turn to how objects are created at a service run by some
principal $P_0$, and
how such objects can be garbage collected when there are no more
outstanding references (codecaps) to those objects.
A client typically needs a codecap with \emph{factory rights}
in order to create new objects.  For example, a directory server
may distribute codecaps with factory rights that can be used
to create new directories.
For convenience, codecaps for factory objects may be available in special
``yellow pages'' directories that are referenced by well-known codecaps.

When principal $P_0$ receives a ``create'' request from a principal $P_1$,
it checks to make sure that $P_1$ has factory rights.  If so, $P_0$ creates
a new object and a corresponding new heritage $h_1$ containing a
certificate $C_1$ that specifies the rights that $P_1$ gets on the object.
Service $P_0$ then sends heritage $h_1$ to $P_1$, which adds its private
key in order to obtain a codecap $c_1$ for the object.

The dual of creation is garbage collection, a difficult problem in
distributed object systems as it is hard to identify which objects are
no longer reachable.
We present a partial solution here.  The idea is that
every object has a ``primary link'' consisting of a directory codecap and
a name.  If there is a codecap for the object stored in the corresponding row
of the directory, then the object persists.
If not, then the object is eventually destroyed.
(An object may optionally support multiple primary links.)

To implement this, the service that provides the object periodically
checks the object's primary link to see if it still points to the object.
If so (or if directory is unavailable), then the object persists.
Otherwise, the object service destroys the object.
If a directory service is discontinued, then there may be a set of
objects that have dangling primary links.  Those primary links are
stored in ``lost+found'' directories.
These directories are
checked and cleaned up manually by an administrator.

\section{Implementation}

Our prototype implementation of codecap authorization is based on
standard X.509 certificates~\cite{RFC-5280} using the widely adopted
OpenSSL\footnote{http://www.openssl.org}
 library and tools.  The X.509 standard defines
several standard fields in certificates including a subject
name, an issuer name, and validity dates.  It enables us to
make use of RSA, DSA, and ECC, with varying key sizes and parameters.
We use established best practices.  Certificates can be either
self-signed, in which case a PKI is not required, or signed by a
common trusted CA.

A codecap heritage is implemented as list of concatenated X.509
proxy certificates as defined in the RFC-3820 standard~\cite{RFC-3820}.
This standard defines the proxyCertInfo
certificate extension containing three fields: path length, policy
language, and policy.  The path length $C.\texttt{pLength}$ is used to
restrict the length a heritage and can be used to implement confinement.
The policy field holds our rights functions $C.\texttt{rights}$ (expressed in
JavaScript), and the policy
language $C.\texttt{pLanguage}$ is set to $anyLanguage$ to indicate
application-specific policies.

Certificate size varies with key size, signature algorithm, and with
the size of the information used to identify subject and issuer.
A certificate may also contain extensions with variable content
length.  A typical PEM encoded certificate combining 2048-bit RSA
public key with SHA-1 and with common extensions like subject key
identifier, authority key identifier, and usage constraints, will
be about 1.2\,KB.  In the more compact DER binary representation,
the same certificate is 0.86\,KB.

Currently we do all communication over SSL, since it is
widely adopted on the Internet for server authentication using X.509
certificates.  By requiring that the optional client authentication
step of the SSL handshake is run, both end-points will mutually
authenticate themselves to each other. The protocol also provides
us with transport level encryption.

After establishing the mutually authenticated SSL connection and
having received the server certificate $C_s$, the client can check
that it is connected to the right service.
The client is free to reject certificates that
do not conform to additional constraints like a valid expiration
date or set usage areas.  If the client accepts the connection, it
will transmit the heritage in combination with its intended request.

Although SSL supports transmission of more than one
certificate from the server to the clients during the handshake,
its intended use is to inform the client about trusted CAs,
and there is no facility for transferring extra certificates
from the client to the server.  Therefore, a codecap containing
multiple certificates cannot be transferred and validated during
the SSL handshake and codecaps must be validated separately.

Having received the client certificate $C_c$, the heritage $h_n$,
and the request $r$, the server will check that:

\begin{itemize}
\item $C_n.\texttt{public} = C_c.\texttt{public}$ (to ensure that
the client is correctly authenticated);

\item for $i = 1,\ldots,n-1$, $C_i.\texttt{subject} = C_{i+1}.\texttt{issuer}$
(to ensure that the heritage is correctly chained);

\item for $i = 1,\ldots,n-1$, $C_i.\texttt{pLength} > C_{i+1}.\texttt{pLength} \geq 0$
(sanity check);

\item the signature of each certificate verifies with the issuer's public key.

\end{itemize}

We have enhanced the Twisted-Python\footnote{\url{http://twistedmatrix.com}}
web-server module with codecap-based authorization.
To transfer the heritage, we extended the commonly used HTTP authentication
mechanism with a codecap credential method.
The client authenticates itself by setting the header field:
\begin{verbatim}
 Authentication: Codecaps <heritage>
\end{verbatim}
where \texttt{<heritage>} is the list of PEM encoded X.509 certificates.
If the header is not provided or the heritage does not validate correctly the server returns a ``401 Unauthorized'' error code and includes the header:
\begin{verbatim}
 WWW-Authenticate: Codecaps realm=<sub> 
\end{verbatim}
where \texttt{<sub>} corresponds to $P_0.\texttt{subject}$ and
is used by the client to identify the correct codecap to use.  If
the same codecap is used to authorize multiple requests, the server
may temporarily store the provided heritage and use a client-side
session cookie to decrease network overhead.

To evaluate the rights function we use the Firefox
SpiderMonkey\footnote{\url{https://developer.mozilla.org/en/SpiderMonkey}}
JavaScript engine. 
When executed, the script is initialized with the following context:
\begin{itemize}
\item \emph{heritage} --- a list of X.509 certificate objects;
\item \emph{idx} --- the position in the heritage list of the certificate currently being evaluated; and
\item \emph{request} --- the client request.
\end{itemize}
Figure~\ref{fig:javascript1} shows a simple rights function that matches the URI of the client's request with
any path restrictions encoded in the common name field of the certificate. 

\begin{figure}
\small
\begin{verbatim}
var allow = heritage[idx].get_subject().CN;
if (request.uri == allow) 1; else 0;
\end{verbatim}
  \caption{A simple JavaScript based rights function}
  \label{fig:javascript1}
\end{figure}

\section{Related Work}

Dennis and Van Horn~\cite{DvH66} first used the term ``capability''
for an unforgeable access token.  Many capability-based systems
have been built, but they usually rely on a trusted runtime environment
in order to prevent forging of capabilities and to mediate communication
of capabilities.  Chaum~\cite{CF78} presents the first cryptographic
approach to capabilities that does not make such an assumption.
The Livermore Network Communication System~\cite{Don81} and the
Amoeba distributed operating system~\cite{MT86} adopted and
improved on this approach~\cite{TMvR86}.  Amoeba also contained a
directory service for capabilities.  However, such capabilities
cannot be confined in any way and rights that can be delegated are
predefined.  Codecaps build on this work, but supports fine-grained
rights delegation through embedded code and supports confinement by
embedding a private key in each codecap.

The capability mechanism proposed by Harnik et al.~\cite{Harnik2011}
uses keyed cryptographic hashes in a way similar to Amoeba and
supports delegation by chaining hashes.  Each entry on the chain
can contain regular expressions to express which rights are being
delegated.  The mechanism is less expensive than our approach,
but does not support rights amplification and cannot be used for
secure abstraction.  The MyProxy service~\cite{Basney2005:MyProxy}
uses X.509 proxy certificates to delegate credentials, but
lacks facilities for including and evaluating complex rights
functions.

Amazon Web Services and Microsoft Azure support capability-like URLs
for use in the
cloud, which contain a query, an expiration time, and a signature.
The query is similar
to the embedded code of rights functions in codecaps.  However, the
URLs cannot be confined or be delegated in a restricted manner, and
the mechanisms do not support rights amplification.

\section{Conclusion}

We have proposed codecaps as a flexible 
way of providing discretionary access control in the cloud.  We are developing this
as part of a sports analytics application that runs on a federated
cloud environment.  
Codecaps are essentially certificate chains corresponding to the chain of delegation,
but has the novel property that they may contain boolean JavaScript function that checks whether a requested operation is allowed.


Using codecaps, we have demonstrated how it is possible to do
fine-grained rights delegation, confinement, and rights amplication
as needed for secure abstraction layers.  We have also shown various
solutions to revocation.  Users can maintain codecaps and facilitate
their delegation using codecap directories.

We have not yet finished the implementation of our codecap-based access
control infrastructure, but soon hope to present experiential
data on its effectiveness.

\bibliographystyle{plain}
\bibliography{all,paper}

\end{document}